\documentclass[12pt,thmsa]{article}
\usepackage{amssymb}
\usepackage{graphicx}

\input{tcilatex}
\begin{document}

\begin{center}
\textbf{\large Tripartite entanglement dynamics for an atom interacting with
nonlinear couplers}

~

\textbf{Mahmoud Abdel-Aty$^{1,2,}$\footnote[5]{Corresponding
author: abdelatyquantum@gmail.com}, M. Sebawe Abdalla$^{3}$ and B.
C. Sanders$^{4}$}

~

{\footnotesize $^{1}$Mathematics Department, Faculty of Science,
Sohag University, 82524 Sohag, Egypt \\[0pt]
$^{2}$Mathematics Department, College of Science, Bahrain
University, 32038 Bahrain \\[0pt]
$^{3}$Mathematics Department, College of Science, King Saud
University, Riyadh 11451, Saudi Arabia }
\\[0pt]
{\footnotesize $^{4}$Institute for Quantum Information Science,
University of Calgary, Calgary, Alberta T2N 1N4, Canada}
\end{center}

\textbf{Abstract:} In this communication we introduce a new model which
represents the interaction between an atom and two fields injected
simultaneously within a cavity including the nonlinear couplers. By using
the canonical transformation the model can be regarded as a generalization
of several well known models. We calculate and discuss entanglement between
the tripartite system of one atom and the two cavity modes. For a short
interaction time, similarities between the behavior based on our solution
compared with the other simulation based on a numerical linear algebra
solution of the original Hamiltonian with truncated Fock bases for each
mode, is shown. For a specific value of the Kerr-like medium defined in this
paper, we find that the entanglement, as measured by concurrence, may
terminate abruptly in a finite time.

~



\section{Introduction}

Recently, the experimental generation of atom--field entangled
states in the mesoscopic field regime has been reported for fields
with an average photon number of a few tens \cite{auf03}. Several
entanglement measures have been studied, for example the von Neumann
reduced entropy~\cite{jv}, the relative entropy of
entanglement~\cite{mp}. Several authors proposed physically
motivated postulates to characterize entanglement
measures~\cite{jv}-\cite{nie00}. Although these postulates vary from
one author to another in the details, however they have in common
that they are based on the concepts of the operational formulation
of quantum mechanics~\cite{kr}. The entanglement properties have
been reported in many different cases \cite{ban05,del98}.

On the other hand, various generalizations of the atom-field interaction
models have been considered. One such generalization is to inject two fields
simultaneously within a high-$Q$ two-mode bichromatic cavity. In this case
the atom interacts with each field individually as well as both fields \cite%
{abd}. In the direction of nonlinear generalization of the atom-field
interaction the influence of nonlinear coupling of a cavity mode with a
Kerr-like medium on the decay of the excited state of the atom was reported
\cite{sho93,bar85}. The nonlinearity can make the dynamics more intricate,
for example with respect to switching, modulation and frequency selection of
radiation in optical communication networks~\cite{jp1,jp2,jp3}. In addition,
the presence of a nonlinear medium is of particular interest as atom-field
dynamics is significantly affected.

\bigskip

This encouraged and stimulated us to investigate a new version of
the atom-field interaction model by including the nonlinear
couplers. The main goal of this work is to investigate the effect of
the nonlinear medium on the phenomenon of entanglement when two
fields simultaneously injected within a high-$Q$ cavity. We
introduce a way to study tripartite entanglement which is an
important issue, by combining two fields and an atom in a double
cavity geometry with a Kerr nonlinearity. We also discuss how the
nonlinear effects increase in importance compared with increasing
atom-field coupling. We show that nonlinear considerations can play
a major role in determining the maximum entangled states of the
system and generating C-Note gate. Particularly, we employ linear
entropy and concurrence to quantify coherence and entanglement,
respectively. Beyond fundamental investigations, these results may
be of benefit in the characterization of atom-field parameters, and
controlling these features may allow for new control techniques for
single and multiple qubit coupling.

\bigskip

The paper is organized as follows: In section 2 we describe the model and
the analytical approach which based on the canonical transformation. In Sec.
3 we briefly comment on the collapse-revival phenomena. In Sec. 4 we
identify the different regimes of entanglement due to the linear entropy and
concurrence. Sec. 5 presents some concluding remarks.

\section{The model and its solution}

We devote this section to introduce our model taking into
consideration the effect of a Kerr-like medium (nonlinear couplers).
For
this reason let us make our starting point the Hamiltonian%
\begin{equation}
\hat{H}=\hat{H}_{\mathrm{f}}+\hat{H}_{\mathrm{a}}+\hat{H}_{\mathrm{af}},
\label{4}
\end{equation}%
where $\hat{H}_{\mathrm{f}}$ comprises a field-field interaction (parametric
down conversion model) in the presence of a nonlinear Kerr medium given by
\begin{equation}
\hat{H}_{\mathrm{f}}=\sum_{i=1}^{2}\hslash \left[ \left( \omega _{i}\hat{a}%
_{i}^{\dagger }\hat{a}_{i}+\chi _{i}\hat{a}_{i}^{\dagger 2}\hat{a}%
_{i}^{2}\right) +\bar{\chi}\hat{a}_{1}^{\dagger }\hat{a}_{1}\hat{a}%
_{2}^{\dagger }\hat{a}_{2}\right] +\hslash \lambda \left( \hat{a}%
_{1}^{\dagger }\hat{a}_{2}+\hat{a}_{2}^{\dagger }\hat{a}_{1}\right) ,
\label{5}
\end{equation}%
where $\chi _{i},i=1,2$ and $\bar{\chi}$ are related to the cubic
susceptibility of the medium, such that $\chi _{i},i=1,2$ \ represent the
self-action for each mode, while $\bar{\chi}$ is related to the cross-action
processes, respectively~\cite{abd1}. The other two parts of the Hamiltonian
consists of the atom-field interaction and the free atomic system that is
\begin{equation}
\hat{H}_{\mathrm{a}}=\frac{\omega _{0}}{2}\hslash \hat{\sigma}_{z},\mathrm{\
\ \ \ \ \ \ }\hat{H}_{\mathrm{af}}=\sum_{i=1}^{2}\hslash \lambda _{i}\left(
\hat{a}_{i}\hat{\sigma}_{+}+\hat{a}_{i}^{\dagger }\hat{\sigma}_{-}\right) .
\label{6}
\end{equation}%
The Hamiltonian (\ref{5}) can be regarded as a generalization of models
considered earlier in the absence of the Kerr-like medium, see for example~%
\cite{abd2}, which concentrate on studying entropy squeezing as well the
degree of entanglement. In order to consider the statistical properties of
the present system we have to find the dynamical operators either by solving
the equations of motion in the Heisenberg picture, or by finding the wave
function in the Schr\"{o}dinger picture.

The latter case will be adopted. To reach our goal let us first introduce
the canonical transformation
\begin{equation}
\left(
\begin{array}{c}
\hat{a}_{1} \\
\hat{a}_{2}%
\end{array}%
\right) =\left(
\begin{array}{cc}
\cos \theta & \sin \theta \\
-\sin \theta & \cos \theta%
\end{array}%
\right) \left(
\begin{array}{c}
\hat{b}_{1} \\
\hat{b}_{2}%
\end{array}%
\right) ,  \label{7}
\end{equation}%
with the properties
\begin{equation}
\hat{a}_{1}^{\dagger }\hat{a}_{1}+\hat{a}_{2}^{\dagger }\hat{a}_{2}=\hat{b}%
_{1}^{\dagger }\hat{b}_{1}+\hat{b}_{2}^{\dagger }\hat{b}_{2},  \label{8}
\end{equation}%
where $[\hat{a}_{i},\hat{a}_{j}^{\dagger }]=[\hat{b}_{i},\hat{b}%
_{j}^{\dagger }]=\delta _{ij}.$ Since the total energy for the system before
the rotation is equal to that after rotation and hence the transformation is
invariant. Thus, with a particular choice of the angle $\theta ,$ the above
transformation would enable us to remove the evanescent waves term from the
Hamiltonian (\ref{5}). This can be achieved if we take
\begin{equation}
\theta =\frac{1}{2}\tan ^{-1}\left( \frac{2\lambda }{\omega _{2}-\omega _{1}}%
\right) .  \label{9}
\end{equation}%
In this case the Hamiltonian (\ref{5}) reduces to
\begin{equation}
\hat{H}=\sum_{i=1}^{2}\hslash \left[ \left( \Omega _{i}\hat{b}_{i}^{\dagger }%
\hat{b}_{i}+\chi \hat{b}_{i}^{\dagger 2}\hat{b}_{i}^{2}\right) +\mu
_{i}\left( \hat{b}_{i}\hat{\sigma}_{+}+\hat{b}_{i}^{\dagger }\hat{\sigma}%
_{-}\right) \right] +\hslash \chi \sum_{i\neq j=1}^{2}\hat{b}_{i}^{\dagger }%
\hat{b}_{i}\hat{b}_{j}^{\dagger }\hat{b}_{j}+\frac{\omega _{0}}{2}\hslash
\hat{\sigma}_{z},  \label{10}
\end{equation}%
where we have assumed the cross-action coupling parameter is equal to twice
the self-action parameter of each mode, so that $\chi _{1}=\chi _{2}=\chi =%
\frac{1}{2}\bar{\chi}$ (codirectional coupler case). Further, we
have also
defined%
\begin{equation}
\Omega _{1}=\left( \omega _{1}\cos ^{2}\theta +\omega _{2}\sin ^{2}\theta
-\lambda \sin 2\theta \right) ^{\frac{1}{2}},\qquad \Omega _{2}=\left(
\omega _{2}\cos ^{2}\theta +\omega _{1}\sin ^{2}\theta +\lambda \sin 2\theta
\right) ^{\frac{1}{2}},  \label{11}
\end{equation}%
and
\begin{equation}
\mu _{1}=\lambda _{1}\cos \theta -\lambda _{2}\sin \theta ,\qquad \mu
_{2}=\lambda _{2}\cos \theta +\lambda _{1}\sin \theta .  \label{12}
\end{equation}%
Finally, let us adjust the coupling parameter $\lambda $ to take the form
\begin{equation}
\lambda =\lambda _{1}\lambda _{2}\left( \frac{\omega _{2}-\omega _{1}}{%
\lambda _{2}^{2}-\lambda _{1}^{2}}\right) ,  \label{13}
\end{equation}%
from which the interaction Hamiltonian $V^{I}(t)$ can be written thus
\begin{equation}
V^{I}(t)=\bar{\mu}\left( \hat{b}_{2}\hat{\sigma}_{+}\otimes \exp \left(
-i[\Delta +2\chi (\hat{b}_{1}^{\dagger }\hat{b}_{1}+\hat{b}_{2}^{\dagger }%
\hat{b}_{2}-1]t\right) +c.c\right) ,  \label{14}
\end{equation}%
where $\bar{\mu}=\sqrt{\lambda _{1}^{2}+\lambda _{2}^{2}}$ and $\Delta
=\Omega _{2}-\omega _{0}.$ Having obtained the interaction picture, we are
therefore in a position to find the explicit solution of the wave function
and consequently the density matrix. After some calculations the density
matrix of the system can be written as

\begin{equation}
\hat{\rho}(t)=U(t)\hat{\rho}(0)U^{\dagger }(t),  \label{dens}
\end{equation}%
where $U(t)$ is a $2\times 2$ matrix representing the time evolution
operator; its elements are
\begin{eqnarray}
U_{11}(t) &=&\left[ \cos \delta _{m_1,m_2}^{+}t+i\frac{\Gamma _{m_1,m_2}}{%
2\delta _{m_1,m_2}^{+}}\sin \delta _{m_1,m_2}^{+}t\right] \exp (-i\Gamma
_{m_1,m_2}t),  \nonumber \\
U_{12}(t) &=&(U_{12})^{\ast }=-i\bar{\mu}\sqrt{m_2+1}\frac{\sin \delta
_{m_1,m_2}^{+}t}{\delta _{m_1,m_2}^{+}}\exp (-i\Gamma _{m_1,m_2}t),
\nonumber \\
U_{22}(t) &=&\left[ \cos \delta _{m_1,m_2}^{-}t-i\frac{\Gamma _{m_1,m_2}}{%
2\delta _{m_1,m_2}^{-}}\sin \delta _{m_1,m_2}^{-}t\right] \exp (i\Gamma
_{m_1,m_2}^{\prime }t),  \label{16}
\end{eqnarray}

\begin{eqnarray}
\ \delta _{m_1,m_2}^{+} &=&\left[ \frac{1}{4}\Gamma _{m_1,m_2}^{2}+\bar{\mu}%
^{2}(m_2+1)\right] ^{\frac{1}{2}},\qquad \delta _{m_1,m_2}^{-}=\left[ \frac{1%
}{4}\Gamma _{m_1,m_2}^{2}+\bar{\mu}^{2}m_2\right] ^{\frac{1}{2}},  \nonumber
\\
\Gamma _{m_1,m_2} &=&\Delta +2\chi (m_1+m_2-1),\qquad \Gamma
_{m_1,m_2}^{\prime }=\Delta +2\chi (m_1+m_2-2).  \label{17}
\end{eqnarray}

\section{Atomic inversion}

For applications in real systems, one can see the atomic inversion of the
two-level system is of particular interest. Therefore, in the present
communication we discuss the time dependence of the two-level-system
observable $\langle \hat{\sigma}_{z}(t)\rangle $, for different values of
the Kerr-like medium. The usual coherent state is used as initial conditions
for the fields. As might be expected, the behavior of the two-level system
changes dramatically depending on the value of the non-linear medium.
\begin{figure}[tbph]
\begin{center}
\includegraphics[width=9cm,height=5cm]{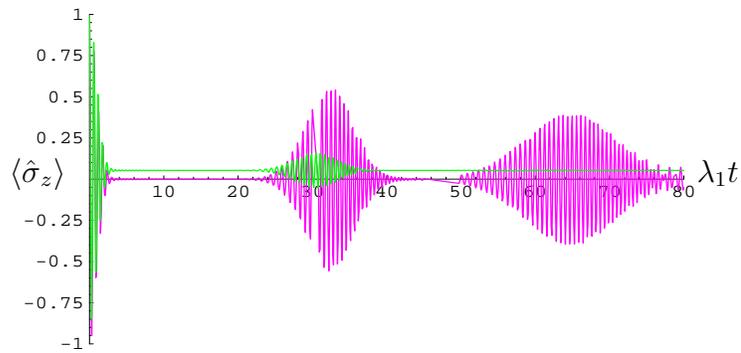} \put(0,70){$\lambda_1 t$ }
\put(-260,70){$\langle\hat{\sigma}_z\rangle $ }
\end{center}
\caption{Atomic inversion as a function of the scaled time $\protect\lambda %
_{1}t$. The parameters are $\protect\lambda _{2}/\protect\lambda _{1}=0.01,%
\protect\omega _{1}/\protect\lambda _{1}=0.2,\protect\omega _{2}/\protect%
\lambda _{1}=0.1,\Delta /\protect\lambda _{1}=0,\protect\chi /\protect%
\lambda _{1}=0$, and $\bar{n}_{1}=\bar{n}_{2}=10$. The initial state is $|%
\protect\alpha _{1},\protect\alpha _{2};e\rangle $, where $\protect\alpha %
_{i}=|\protect\alpha _{i}|e^{i\protect\phi },$ $\protect\chi /\protect%
\lambda _{1}=0.001$ (solid curve) and $\protect\chi /\protect\lambda %
_{1}=0.1 $ for (dots). }
\label{at1}
\end{figure}

In Fig. (\ref{at1}) we have plotted the atomic inversion as a function of
the scaled time $\lambda _{1}t$ for different values of the Kerr-like medium
and the phase of coherent states is taken to be zero i.e. $\phi =0$. It has
been observed that for a small value of the Kerr-like medium such as $\chi
/\lambda _{1}=0.001,$ the system shows period of collapse after onset of the
interaction. This is followed with a long period of the revival. Here we may
point out that comparing with the first period of collapse the amplitude of
the oscillations for the second period is decreased, see fig.(1) (solid
curve). This means that as the time of the interaction increases the
amplitude of the oscillations decreases. Increasing the value of the
Kerr-like medium leads to increase in the period of collapses with
decreasing in the size of the amplitude of each period, see fig.(1) (dot
curve). It is also noted that as a special case of the model when $\lambda
_{2}=\chi _{2}=0$, the general behavior of the atomic inversion coincides
with that of the well known JCM in presence of the Kerr-like medium \cite%
{buz90}.

We have verified the behavior obtained in the\ numerical simulations based
on the final equations and the numerical solutions of the original
Hamiltonian i.e. one simulation simply based on the final equations and the
other simulation based on a numerical linear algebra solution of the
original Hamiltonian with truncated Fock bases for each mode (see Fig. \ref%
{at2}). Provided the ratio $\lambda _{2}/\lambda _{1}$ is made small enough,
we have excellent agreement for the short time scales. In the meantime, the
value $\lambda _{2}/\lambda _{1}=0.01$ gives quantitative agreement over
some periods of oscillation.

\begin{figure}[tbph]
\begin{center}
\includegraphics[width=9cm,height=5cm]{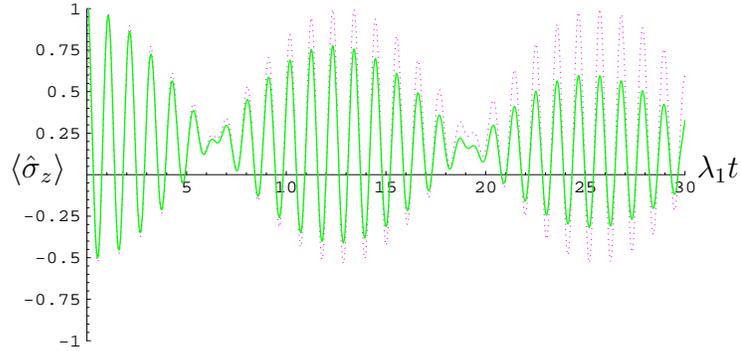} \put(0,70){$\lambda_1 t$ }
\put(-260,70){$\langle\hat{\sigma}_z\rangle $ }
\end{center}
\caption{Atomic inversion as a function of the scaled time $\protect\lambda %
_{1}t$. The parameters are $\protect\lambda _{2}/\protect\lambda _{1}=0.01,%
\protect\omega _{1}/\protect\lambda _{1}=0.2,$ $\protect\chi /\protect%
\lambda _{1}=0.01$ $\protect\omega _{2}/\protect\lambda _{1}=0.1,$ $\Delta /%
\protect\lambda _{1}=0,$ $\protect\chi /\protect\lambda _{1}=0.01$. The
initial state is $|5,6;e\rangle $, where dotted curve obtained using the
final equations and solid curve obtained using linear algebra solution of
the original Hamiltonian.}
\label{at2}
\end{figure}
The revival time for the present system is calculated to give us
\begin{equation}
t_{R}=\frac{2n\pi }{\lambda \left( \Omega _{\bar{n}}-\Omega _{\bar{n}%
-1}\right) },  \label{18}
\end{equation}%
where $\Omega _{\bar{n}}$ is the Rabi frequency, such that $\Omega _{\bar{n}%
}=\delta _{\overline{m}_{1},\overline{m}_{2}}^{+}\ $and $n$ is an integer,%
\begin{eqnarray}
\lambda _{1}t_{R} &=&\frac{2n\pi }{\lambda \Delta \chi ^{\prime }+\chi
^{\prime 2}(4\bar{n}-3)-\overline{\mu }^{\prime 2}}\left\{ \frac{1}{2}\sqrt{%
(2\chi ^{\prime }(2\bar{n}-1)+\Delta ^{\prime })^{2}+\overline{\mu }^{\prime
2}(\bar{n}+1)}\right.  \nonumber \\
&&+\left. \frac{1}{2}\sqrt{(2\chi ^{\prime }(2\bar{n}-2)+\Delta ^{\prime
})^{2}+\overline{\mu }^{\prime 2}\bar{n}}\right\} ,  \label{19}
\end{eqnarray}%
where $\chi ^{\prime }=\chi /\lambda _{1},$ $\overline{\mu }^{\prime }=%
\overline{\mu }^{\prime }/\lambda _{1},$ $\Delta ^{\prime }=\Delta /\lambda
_{1}.$ In the following section we turn our attention to discuss the effect
of the Kerr-like medium on the phenomenon of entanglement.

\section{Entanglement}

In this section we shall concentrate on the discussion of the entanglement
where the total state vector can not be written precisely as the product of
a time-dependent atomic and field component vector.

\subsection{Coherence loss}

Distillable entanglement is a critical resource for quantum information. In
our case we are considering transduction of quantum information between
fields and atoms so the degree of entanglement between the atom and the
field is important to assess its use in quantum information. Furthermore
there are three entities: two field modes and one atom, and the entanglement
between any single entity and the other two can be used so we assess the
inherent resource in our system by calculating the tripartite entanglement
in the combined system.

Here we use the idempotency defect \cite{zur93}, defined by linear entropy,
as a measure of the degree of mixture for a state $\hat{\rho}_{a}(t).$ This
version of entropy makes the problem tractable \cite{ber03} and therefore we
use this idempotency defect as a measure of coherence loss. This is given by
\begin{equation}
\mathcal{E}_{t}^{(a)}=\mathrm{Tr}\big[\hat{\rho}_{a}(t)\left( 1-\hat{\rho}%
_{a}(t)\right) \big],  \label{lin}
\end{equation}%
where $\mathcal{E}_{t}^{(a)}$ has a zero value for a pure state and $1$ for
a completely mixed state.

\begin{figure}[tbph]
\begin{center}
\includegraphics[width=8cm,height=4cm]{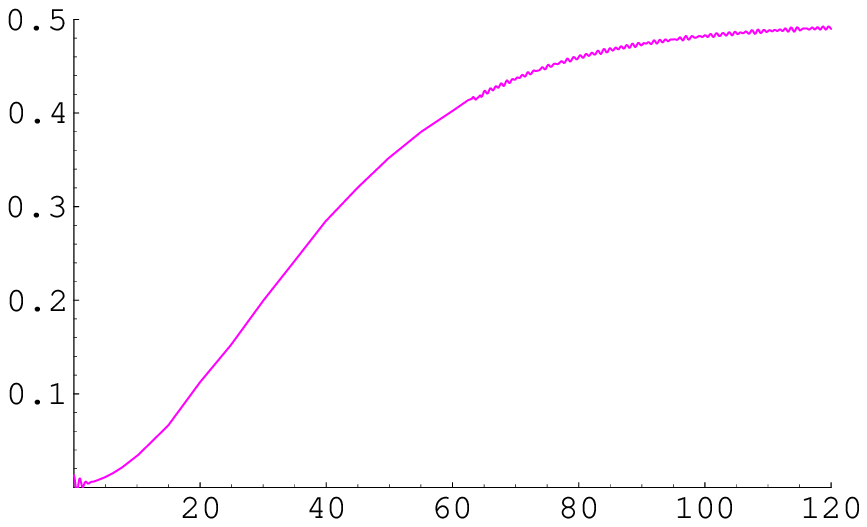} \put(-105,-10){$\lambda_1
t $ } \put(-200,100){(a) } \put(-245,70){$\mathcal{E}_{t}^{(a)}$ } %
\includegraphics[width=8cm,height=4cm]{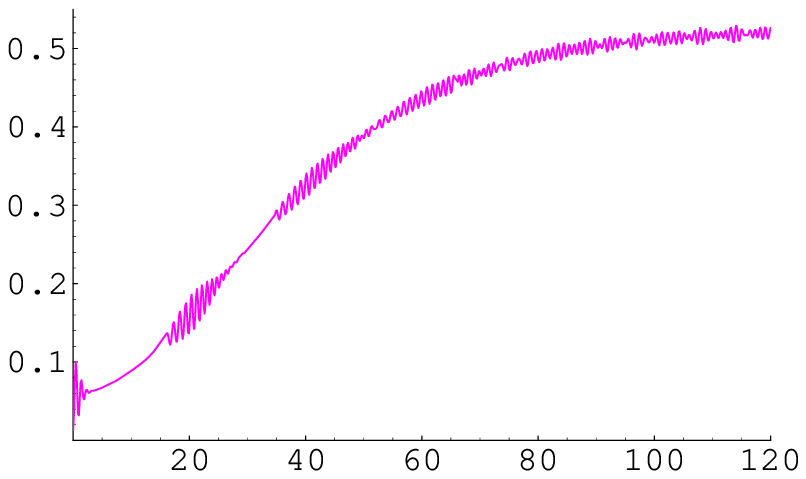} \put(-105,-10){$\lambda_1
t $ } \put(-200,100){(b) }
\end{center}
\caption{Linear entropy as a function of the scaled time $\protect\lambda %
_1t $. The parameters are $\protect\lambda _2/\protect\lambda _1=0.1,\protect%
\omega _1/\protect\lambda _1=0.2,\protect\omega _2/\protect\lambda %
_1=0.1,\Delta /\protect\lambda _1=0,\protect\chi /\protect\lambda _1=0$, and
$\bar{n}_1=\bar{n}_2=10$. The initial state is $\frac{1}{\protect\sqrt{2}}%
(|e\rangle +|g\rangle )\otimes |\protect\alpha _1,\protect\alpha _2\rangle $%
, where (a) $\protect\chi /\protect\lambda _1=0.001$ and (b) $\protect\chi /%
\protect\lambda _1=0.01$. }
\label{lin1}
\end{figure}

In Fig. (\ref{lin1}) we have plotted the idempotency defect as a function of
the scaled time for different values of the Kerr-like medium. It is shown
that the asymptotic value of the linear entropy is obtained when the time is
increased. Also the value of $\mathcal{E}_{t}^{(a)}$ increases as the time
increases in addition to the appearance of irregular fluctuations at
discrete period of time for a large value of the Kerr-like medium, see
fig.(3b). Of course, there are some differences between the two cases (small
and large values of the Kerr-like medium) in the amplitudes but the general
behavior is the same. This may be thought to arise from the asymptotic
limits which have been observed in both Figs. \ref{lin1}(a,b).

\subsection{Tripartite Quantum States}

Now we would like to discuss an example of entanglement between the two
cavity fields and atom. In the case of pure state, if the density matrix
obtained from the partial trace over other subsystems is not pure the state
is entangled. Consequently, for the pure state of a bipartite system,
entropy of the density matrix associated with either of the two subsystems
is a good measure of entanglement. In the mixed state case, the entanglement
can be quantified by the quantity $C(\rho )$ which is known in literature as
concurrence. Quite recently, some approaches have been reported for the
determination of entanglement in experiment \cite{wal07,wal06,min07,min05}.
The most remarkable are the new formulation of concurrence \cite{min05} in
terms of copies of the state which led to the first direct experimental
evaluation of entanglement and some analogous contributions \cite{aol06} to
multipartite concurrence.

For the density matrix $\hat{\rho}(t),$ which represents the state of a
bipartite system, concurrence is defined as \cite{woo98}
\begin{equation}
C(\hat{\rho})=\max \{0,\mathit{\lambda }_{1}-\mathit{\lambda }_{2}-\mathit{%
\lambda }_{3}-\mathit{\lambda }_{4}\},  \label{con}
\end{equation}%
where the $\mathit{\lambda }_{i}$ are the non-negative eigenvalues, in
decreasing order ($\mathit{\lambda }_{1}\geq \mathit{\lambda }_{2}\geq
\mathit{\lambda }_{3}\geq \mathit{\lambda }_{4}$), of the Hermitian matrix
\begin{equation}
\widehat{\Upsilon }\equiv \sqrt{\sqrt{\hat{\rho}}\widetilde{\rho }\sqrt{\hat{%
\rho}}},\,\widetilde{\rho }=\left( \widehat{\sigma }_{y}\otimes \widehat{%
\sigma }_{y}\right) \hat{\rho}^{\ast }\left( \widehat{\sigma }_{y}\otimes
\widehat{\sigma }_{y}\right) .
\end{equation}%
Here, $\hat{\rho}^{\ast }$ represents the complex conjugate of the density
matrix $\hat{\rho}$ when it is expressed in a fixed basis and $\widehat{%
\sigma }_{y}$ represents the Pauli matrix in the same basis. The function $C(%
\hat{\rho})$ ranges from $0$ for a separable state to $1$ for a maximum
entanglement.

We first investigate the quantum correlation between the atom and cavity
modes. If we consider the initial state of the system is given by $\rho
(0)=\gamma |0,1;e\rangle \langle 0,1;e|+\beta |0,1;g\rangle \langle 0,1;g|,$
$\gamma +\beta =1,$ which means that the atom starts from a mixed state and
the field starts from $|0,1\rangle $ state i.e. the vacuum for the first
mode and one photon in the other mode. If we deal with the two cavity modes
as system $A$, and the atom as system $B$, then $\rho (t)$ in equation (\ref%
{dens}) can be thought of as the density operator of a two-qubit mixed
state. In the basis $|1\rangle \equiv |01\rangle \otimes |e\rangle $, $%
|2\rangle \equiv |01\rangle \otimes |g\rangle $, $|3\rangle \equiv
|00\rangle \otimes |e\rangle $, $|4\rangle \equiv |02\rangle \otimes
|g\rangle $, the density matrix can be written as \cite{san00}

\begin{equation}
\rho (t)=\left(
\begin{array}{c}
\rho _{11} \\
0 \\
0 \\
\rho _{41}%
\end{array}%
\begin{array}{c}
0 \\
\rho _{22} \\
\rho _{32} \\
0%
\end{array}%
\begin{array}{c}
0 \\
\rho _{23} \\
\rho _{33} \\
0%
\end{array}%
\begin{array}{c}
\rho _{14} \\
0 \\
0 \\
\rho _{44}%
\end{array}%
\right) .  \label{den}
\end{equation}%
where $\rho _{ij}=\langle i|\rho (t)|j\rangle .$

The explicit expression of the concurrence describing the entanglement
between the system $A$ and system $B$ can be found using equations (\ref{con}%
) and (\ref{den}) as\bigskip\
\begin{equation}
C(\rho )=2\max \left\{ 0,|\rho _{23}|-\sqrt{\rho _{11}\rho _{44}},|\rho
_{14}|-\sqrt{\rho _{22}\rho _{33}}\right\} ,
\end{equation}

\begin{figure}[tbph]
\begin{center}
\includegraphics[width=8cm,height=4cm]{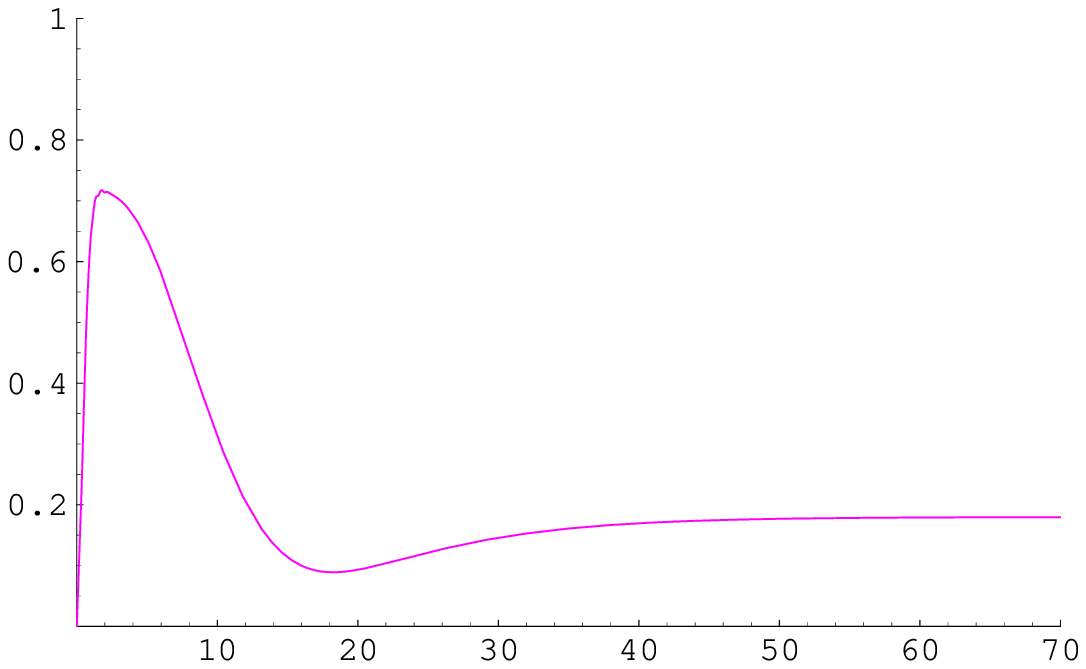} \put(-105,-10){$\lambda_1 t$
} \put(-200,100){(a) } \put(-245,74){$C(\hat{\rho})$ } %
\includegraphics[width=8cm,height=4cm]{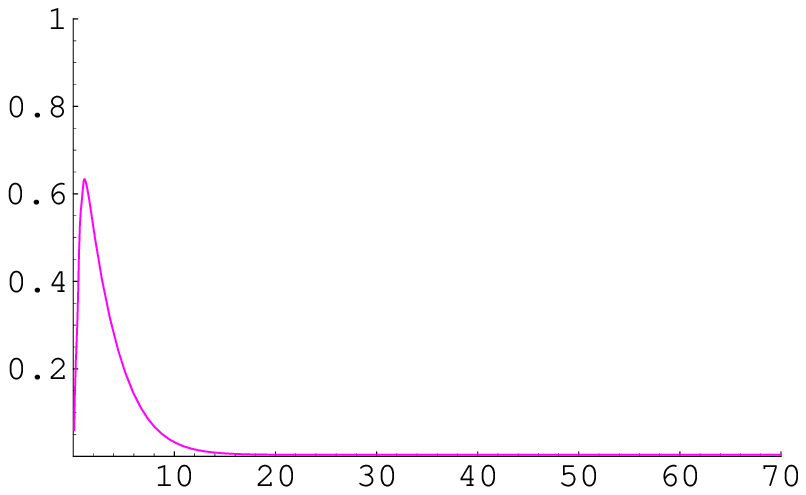} \put(-105,-10){$\lambda_1 t$
} \put(-200,100){(b) }
\end{center}
\caption{Concurrence as a function of the scaled time $\protect\lambda _1t$.
The parameters are $\protect\lambda _2/\protect\lambda _1=0.1,\protect\omega %
_1/\protect\lambda _1=0.2,\protect\omega _2/\protect\lambda _1=0.1,\Delta /%
\protect\lambda _1=0,\protect\chi /\protect\lambda _1=0$. The initial state
is $\protect\rho (0)=\frac{1}{2}(|e\rangle \langle e|+|g\rangle \langle
g|)\otimes |0,1\rangle \langle 0,1|$, where (a) $\protect\chi /\protect%
\lambda _1=0.001$ and (b) $\protect\chi /\protect\lambda _1=0.01$. }
\end{figure}

The graphs of the concurrence as a function of time for the two dynamical
regimes are displayed in Fig. 4. Keeping the value of the Kerr-like medium
small enough, the asymptotic value of the concurrence is not null, since the
global system evolves to a classically correlated state. But once the $\chi
/\lambda _{1}$ is increased we see that the concurrence vanishes in the
asymptotic limit. With increasing $\chi /\lambda _{1}$ further, we note that
the decay of the concurrence is more rapid than the corresponding decay in
the weak nonlinear regime.

If the initial state has been considered such as $\gamma =1,$ then the
sudden death of entanglement will be obtained if $\rho _{11}\rho _{44}=0.$
This means that the sudden death time is given by
\begin{equation}
\lambda _{1}t_{d}=\frac{1}{2\chi ^{\prime }-\Delta ^{\prime }}\cos
^{-1}\left( \frac{2\left( \Delta ^{\prime }-2\chi ^{\prime }\right) ^{2}}{%
\left( \Delta ^{\prime }-2\chi ^{\prime }\right) ^{2}-8\left( 1+\lambda
_{2}^{\prime 2}\right) }\right) .
\end{equation}%
In the limit that the fields decouple from the atom, it is shown that, one
may just entangling the cavity fields along the lines which has been
considered in Ref. \cite{san92}.

In quantum computation operations are performed by means of single-qubit and
multiple-qubit quantum logic gates. In what follows we consider universal
quantum logic gate based on two electromagnetic field modes of a cavity.
From equation (\ref{dens}), we can get various time evolutions of the
present system. If the initial state is taken to be $|e\rangle ,$ and we set

\begin{equation}
\chi =\frac{\lambda _{1}}{2}\sqrt{4-\overline{\mu }^{2}},
\end{equation}%
then, to obtain the C-NOT gate, we need the unitary operation with $\chi =%
\frac{\lambda _{1}}{2}\sqrt{4-\overline{\mu }^{2}},$ $\Delta =0,$ and
interaction time $\lambda _{1}t=2n\pi $, ($n=1,2,3,...$) which gives

\begin{eqnarray}
|e\rangle _{a}|0,0\rangle _{f} &\mapsto &|e\rangle _{a}|0,0\rangle
_{f}\mapsto |e\rangle _{a}|0,0\rangle _{f}\mapsto |e\rangle _{a}|0,0\rangle
_{f},  \nonumber \\
|g\rangle _{a}|0,1\rangle _{f} &\mapsto &|g\rangle _{a}|0,1\rangle
_{f}\mapsto |g\rangle _{a}|0,1\rangle _{f}\mapsto |g\rangle _{a}|0,1\rangle
_{f},
\end{eqnarray}%
i.e. using a specific values of the Kerr-like medium in the present model,
one can be able to implement a C-NOT gate.

\section{Conclusion}

In this paper we have discussed a new Hamiltonian class which
describes the the interaction between a single two-level atom and
bimodal cavity field taking into account an optical Kerr
nonlinearity. We present an analytically solution and numerical
investigation of the atomic inversion, coherence loss and
entanglement. Entanglement of the tripartite system of one atom
and the two cavity modes has been discussed and the sudden death
of entanglement is shown. Also, we proposed a scheme for quantum
computing (C-NOT gate), which is realized by a nonlinear
interaction in the QED cavity. For our model to be useful for
experiments, we would need to include cavity losses, fluorescence,
and atomic motion. These extensions to our model could be
accomplished by standard master equation methods, and such an
analysis is reserved for future work. Our purpose here has been to
develop a Hamiltonian that includes many of the iconic
Hamiltonians used in quantum optics as special cases so that three
coupled systems are used, which opens up investigations from
bipartite to tripartite entanglement. Cavity quantum
electrodynamics with one atom and two fields, including a
nonlinear
interaction, creates rich, interesting dynamics as shown here.%
\[
\]%
\textbf{Acknowledgements}:

B C Sanders$^1$, M. S. Abdalla$^2$ and M. Abdel-Aty$^3$ are
grateful for the financial support from iCORE and a CIFAR
Associateship$^1$, the project Math 2008/132 of the research
center, College of Science, King Saud University$^2$ and the
project No. 11/2008 Bahrain University$^3$.

\end{document}